\documentclass{article}
\pdfoutput=1
\usepackage{enumerate}
\usepackage{amsfonts}
\usepackage{amsmath}
\usepackage{comment}
\usepackage{textcomp}
\usepackage{amsthm, tikz}
\usepackage{amssymb}
\usepackage{color}
\usepackage{graphicx}
\usepackage{bbm}
\usepackage[matrix,arrow,curve]{xy}
\usepackage{array}
\usepackage{mathtools}

\def\be{\begin{eqnarray}}
\def\ee{\end{eqnarray}}
\def\nn{\nonumber}

\usepackage{tikz}
\usepackage{braids}

\def\l[{\phantom.[}

\hoffset= - 1.0in         

\begin{document}

\title{\vspace{.1cm}{\Large {\bf Racah matrices and hidden integrability in evolution of knots}\vspace{.2cm}}
\author{
{\bf A.Mironov$^{a,b,c,d}$}\footnote{mironov@lpi.ru; mironov@itep.ru},
\ {\bf A.Morozov$^{b,c,d}$}\thanks{morozov@itep.ru},
\ {\bf An.Morozov$^{c,d,e}$}\footnote{andrey.morozov@itep.ru},
\ \ and
 \ {\bf A.Sleptsov$^{b,c,d,e}$}\thanks{sleptsov@itep.ru}}
\date{ }
}

\maketitle

\vspace{-5.5cm}

\begin{center}
\hfill FIAN/TD-12/16\\
\hfill IITP/TH-08/16\\
\hfill ITEP/TH-10/16
\end{center}

\vspace{3.3cm}

\begin{center}
$^a$ {\small {\it Lebedev Physics Institute, Moscow 119991, Russia}}\\
$^b$ {\small {\it ITEP, Moscow 117218, Russia}}\\
$^c$ {\small {\it Institute for Information Transmission Problems, Moscow 127994, Russia}}\\
$^d$ {\small {\it National Research Nuclear University MEPhI, Moscow 115409, Russia }}\\
$^e$ {\small {\it Laboratory of Quantum Topology, Chelyabinsk State University, Chelyabinsk 454001, Russia }}

\end{center}

\vspace{.5cm}

\begin{abstract}
We construct a general procedure to extract the exclusive Racah matrices $S$ and $\bar S$ from the inclusive 3-strand mixing matrices by the evolution method and apply it to the first simple representations $R =[1]$, $[2]$, $[3]$ and $[2,2]$.
The matrices $S$ and $\bar S$ relate respectively the maps $(R\otimes R)\otimes \bar R\longrightarrow R$
with $R\otimes (R \otimes \bar R) \longrightarrow R$
and $(R\otimes \bar R) \otimes R \longrightarrow R$
with $R\otimes (\bar R \otimes R) \longrightarrow R$.
They are building blocks for the colored HOMFLY polynomials
of arbitrary arborescent (double fat) knots.
Remarkably, the calculation realizes an unexpected
integrability property underlying the evolution matrices.
\end{abstract}

\vspace{.5cm}

\section{Introduction}

Evaluation of colored link polynomials \cite{knotpols,con}
(Wilson loop averages in Chern-Simons theory \cite{CS} and their stringy generalizations)
is a hard old problem,
where some advance has become possible only recently,
after development of new theoretical methods and increase of the computer power.
It now attracts a lot of attention, also because the resulting polynomials
are the closest relatives of conformal blocks and one expects them to have
even more interesting and intriguing properties.
Approaches to the problem can be very different, still the main advances
so far come from the modern version \cite{modRT1}-\cite{modRT2} of the
Reshetikhin-Turaev (RT) method \cite{RT},
which reduces it to study of the quantum ${\cal R}$-matrices
in the Tanaka-Krein (representation) space and the Racah matrices.
For very promising alternative approaches, see \cite{Che,Den,GLL}.

The difficult part of RT approach is evaluation of the Racah matrices $U$
which relate the intertwiners:
\be
U_{R_1,R_2,R_3}^{R_4}: \ \ \ \ \
\Big\{ (R_1\otimes R_2)\otimes R_3 \longrightarrow R_4\Big\} \ \ \longrightarrow \ \
\Big\{ R_1\otimes (R_2\otimes R_3) \longrightarrow R_4\Big\}
\ee
i.e. describe deviations from the associativity in the product of representations.
Actually, they describe a map from the space of representations $Y$ in the product
$R_1\otimes R_2=\oplus Y_{12}$ into that in $R_2\otimes R_3 = \oplus Y_{23}$.
In the simplest knot theory applications, one needs two types of such matrices:
\be
{\rm inclusive}: \ \ \ {\cal U}_Q\ \ {\rm with}\ \ R_1=R_2=R_3=R, \ \ \ \ R_4 = Q \in R^{\otimes 3}
\label{inc}
\ee
and
\be
{\rm exclusive}:\ \ \ S \  \ {\rm with} \ \ R_1=R_2=R_4=R, \ \ \ \ R_3 = \bar R \nn \\
\ \ \ \ \ \ {\rm or} \ \ \ \ \ \
\bar S \ \ {\rm with} \ \ R_1=R_3=R_4=R, \ \ \ \ R_2 = \bar R
\label{exc}
\ee

The inclusive (the term refers to arbitrariness of the final representation $Q\in R^{\otimes 3}$)
matrices ${\cal U}_Q$ define the $R$-colored HOMFLY polynomials for arbitrary 3-strand braids
${\cal L} = (m_1,n_1|m_2,n_2|\ldots)$ as \cite{MMMkn12},
\be
H_R^{(m_1,n_1|m_2,n_2|\ldots)}(A,q) =
\sum_{Q\in R^{\otimes 3}} \frac{d_Q}{d_R}\cdot
{\rm Tr}_Q \!\Big({\cal R}_Q^{m_1} {\cal U}_Q {\cal R}_Q^{n_1} {\cal U_Q}^\dagger
{\cal R}_Q^{m_2} {\cal U}_Q {\cal R}_Q^{n_2} {\cal U_Q}^\dagger\ldots \Big)
\label{3str}
\ee
where $d_R$ is the quantum dimension of representation $R$ for the Lie algebra $sl_N$,
expressed through the variable $A=q^N$, and ${\cal R}_Q$ is a diagonal matrix
with the entries
\be
\lambda_Y = \epsilon_Y q^{\varkappa_Y}
\label{evY}
\ee
for $Y\in R^{\otimes 2}$.
Here $\varkappa_Y = \sum_{(i,j)\in Y} (i-j)$ is the value of Casimir operator in the representation $Y$,
while $\epsilon_Y = \pm 1$ depending on whether $Y$ belongs to the symmetric or antisymmetric
square of $R$.
For other simple Lie algebras similar formulas exist, see \cite{MMuniv} for a short survey.

The exclusive matrices $S$ and $\bar S$, where only $R$ is picked up in the "final state" of the product
$R\otimes R\otimes \bar R$, define \cite{MMMRS,MMMRSS} the building blocks ("fingers") for $R$-colored
HOMFLY for arbitrary arborescent (double-fat) knots \cite{con,arbor}
${\cal K} = \{F^{I,k_I}\}$:
\be
H_R^{\{F_I\}} = \sum_{X_I \in R\otimes R \ {\rm or}\ R\otimes \bar R} \prod_{I,J} P_{X_I,X_J}
\prod_{k_I} F^{\{I,k_I\}}_{X_I}
\label{arbor}
\ee
where the propagators $P_{X'X''}$ connecting the vertices $I$ are just the matrices $S_{\bar X'X''}$
or $\bar S_{\bar X'\bar X''}$ (bars refer to the antiparallel rather than parallel double lines,
the two parallel vertices never being connected), while the fingers attached to the vertices
are arbitrary matrix elements of the type
\be
F_X =
\Big(\ldots S\bar {\cal R}^{l_3} S{\cal R}^{l_2} S^\dagger \bar{\cal R}^{l_1}\bar S\Big)_{\emptyset X}
\ee

\section{State-of-the-art Racah matrices}

While the matrices ${\cal U}_Q$ and $S,\bar S$ are well known for symmetric (and antisymmetric)
representations $R=[r]$ (and $R=[1^r]$), \cite{Racah,MMSpret} their evaluation for all other $R$ remains a big problem.

For ${\cal U}_Q$, the best at the moment is the highest weight method of \cite{MMMkn12},
it allowed us to find them for $R=[21]$ in \cite{MMMS21}, for $R=[31]$ in \cite{MMMS31} and for $R=[22]$ in
\cite{MMMS22}. The method is very straightforward but extremely tedious, especially for non-rectangular
diagrams $R$ like $[21]$ and $[31]$ (for the rectangular $R$ there are no multiplicities, and things
are considerably simpler, almost as simple as they are for the symmetric representations).
There can be further advances related to the eigenvalue hypothesis \cite{IMMMev} and to
the quantum Vandermonde method mentioned in \cite{MMMS31}.

Still, at the moment the inclusive matrices ${\cal U}_Q$ are not available in general form
(for arbitrary $r$) even in the symmetric case $R=[r]$.

The exclusive $S$ and $\bar S$ are known for arbitrary $R=[1^r]$ \cite{NRZ,MMSpret} and actually look like
straightforward quantization and extension of the classical formulas cited in \cite{LL}.
However, the highest weight method is now difficult, because the conjugate representations
and thus the Racah matrices depend on $N$, thus, one needs to do calculations for various $N$
and reconstruct $N$ dependence from the collection of the answers.
A hard effort in \cite{GJ} allowed them to find these matrices by brute force for $R=[21]$,
but things get very difficult beyond it.

The purpose of this paper is to suggest a knot theory trick that allows one to extract $S$ and $\bar S$
from ${\cal U}_Q$.
This is conceptually strange to find $S$ and $\bar S$ from ${\cal U}_Q$, because the exclusive matrices are in certain sense simpler than the inclusive ones, at the same time it allows us to get $S$ and $\bar S$ for $R=[31]$ and $[22]$
from the 3-strand calculus advance in \cite{MMMS31} and \cite{MMMS22} right now,
without developing any special new technique.
We also reproduce in this simple way the excruciating result of \cite{GJ} for $R=[21]$.

Throughout the text we use the notation:
\be
A=q^N,\ \ \ \ \ \ \{x\}\equiv x-{1\over x},\ \ \ \ \ \ \ D_k={\{Aq^k\}\over \{q\}},\ \ \ \ \ \ \ t=A^{-1}\{q\}
\ee
Also a word of precaution is necessary: in the paper, we use the term "orthogonal matrix" for matrices from the group $O(N)$, they usually have the determinant equal to -1.

\section{The trick}

What we suggest is to extract $S$ and $\bar S$ from the intersection of 3-strand braid and arborescent worlds.
If there are many enough knots which are simultaneously 3-strand and arborescent,
one can extract these matrices from  (\ref{arbor}), where the l.h.s. is calculated with the
help of  (\ref{3str}).

This is especially simple for $S$, because there is a two-parametric family (even two),
which is simultaneously 3-strand braid $(m,-1|\pm n,-1)$ and pretzel $Pr(m,n,\pm\bar 2)$.
For this pretzel family, (\ref{arbor}) simplifies greatly:
\be
H_R^{Pr(m,n,\overline{\pm 2})} =
d_R\sum_{\bar X\in R\otimes \bar R}
\frac{(ST^mS^\dagger)_{\emptyset \bar X}  (ST^nS^\dagger)_{\emptyset \bar X} (\bar S \bar T^{\pm 2} \bar S)_{\emptyset \bar X}}
{S_{\emptyset \bar X}}
=  d_R^{-1} \!\!\!\! \sum_{\stackrel{\bar X\in R\otimes \bar R}{ Y,Z\in R\otimes R}}
\sqrt{ d_Yd_Z} K_{\bar X} S_{\bar X Y}S_{\bar X Z}\cdot \lambda_Y^m\lambda_Z^n
\label{Hpre}
\ee
where  $\lambda_Y$ is the eigenvalue (\ref{evY}), the square
$S^2_{\emptyset Y} =d_Y/d_R^2$ and
\be
K_{\bar X} = d_Rd_X^{-1/2}(\bar S \bar T^{\pm 2} \bar S)_{\emptyset \bar X}
\label{KX}
\ee

The pretzel formula (\ref{Hpre}) should be compared with the answer (\ref{3str})
\be
H_R^{(m,-1|\pm n,-1)} = \sum_{Y,Z\in R^{\otimes 2}} h_{YZ}\cdot \lambda_Y^m \lambda_Z^n
\ee
which is the usual evolution formula, of the type considered in sec.5 of \cite{MMMS31}.
We present it in the next section \ref{hYZ}.
Comparing gives:
\be\label{12}
\boxed{
\sum_X  K_{\bar X} S_{\bar X Y} S_{\bar XZ} = \frac{d_R}{\sqrt{d_Y d_Z}}\cdot h_{YZ}
 = \mathfrak{h}_{YZ}
}
\ee
i.e. $F_{\bar X}$ are the eigenvalues of the matrix at the r.h.s.
(for which we introduce a special notation $\mathfrak{h}$), while our needed
{\bf $S_{\bar XY}$ is the orthogonal diagonalizing matrix} (i.e. the matrix made from the
normalized eigenvectors).
This provides the manifest expressions for $S$ in sec.\ref{hYZ} below.
We should stress that this calculation works this simple way only in the case without multiplicities. Otherwise, there are a few additional complications: size of the matrix $\mathfrak{h}_{YZ}$ is smaller than that of the matrix $S$ (due to additional indices, see \cite{MMMRSS}); there is also a small sign ambiguity depending on the choice of the basis vectors,
which is well-known \cite{MMMRS,MMMRSS} to be significant
for knot polynomial calculus in (non-symmetric) representations
with multiplicities, etc.

\bigskip

To evaluate $\bar S$, just the same trick would require a pretzel family $Pr(\bar m,\bar n, \ldots)$
with two barred (antiparallel) parameters.
Unfortunately there are none of them, which are 3-strand braids.
However, after one knows $S$, one can actually take {\it any} arborescent family which depends on $\bar S$
in a simple way (linearly or quadratically), while can have quite a complicated dependence on $S$.
Actually, there are many choices of this type at the intersection of 3-strand braids and arborescent knots.
Though technically it is equally simple, and leads to the answer,
this trick is somewhat less elegant than the one we use for $S$.
Therefore, for $\bar S$, we use an
alternative way: just to extract it from the known $S$ by making use of the relation (63)
from \cite{MMMRS}:
\be
\boxed{
\bar S =   \bar T^{-1} S T^{-1} S^\dagger \bar T^{-1}
}
\label{bSfromS}
\ee

In the remaining part of the paper we apply these ideas to find $S$ and $\bar S$ in some simple examples and evaluate the colored HOMFLY for the arborescent knots.

\section{Matrix $S$ from the evolution for first representations
\label{hYZ}}

\subsection{Fundamental representation $R=[1]$}

In this case  $Y,Z\in [1]^2 = [11]\oplus [2]$, and the matrix $h_{YZ}$ is $2\times 2$.
Also, dimensions are $d_{[1]}=\frac{\{A\}}{\{q\}}$,
$d_{[11]} = d_{[1]}\cdot\frac{\{A/q\}}{\{q^2\}}$, $d_{[2]} = d_{[1]}\cdot\frac{\{Aq\}}{\{q^2\}}$
and eigenvalues $\lambda_{[11]}=-\frac{1}{qA}$, $\lambda_{[2]}=\frac{q}{A}$.
It is easy to evaluate
\be
\mathfrak{h}^{[1]}_{YZ} = \frac{d_{[1]}}{\sqrt{d_Yd_Z}}\cdot h^{[1]}_{YZ}
\ee
from the 3-strand formula (\ref{3str}), where the only non-trivial mixing matrix is
$\ {\cal U}_{[21]} = \frac{1}{[2]}\left(\begin{array}{cc}1 & \sqrt{[3]} \\ \sqrt{[3]} & -1 \end{array}\right)$:
\be
\mathfrak{h}^{[1]} = \frac{A^2}{[2]\{q\}} \left(\begin{array}{c|cc}
& [11] & [2] \\
\hline
&&\\
\l[11] & q^{-1}A-( q^3-q^{-1}+q^{-3})A^{-1} & \sqrt{\{Aq\}\{A/q\}} \\ && \\
\l[2] & \sqrt{\{Aq\}\{A/q\}} &   qA-(q^3-q+q^{-3})A^{-1}
\end{array}\right)
\ee
The first line and row correspond to representation $[11]$, the second ones to $[2]$.
This symmetric matrix is diagonalized by the orthogonal  matrix
\be
S^{[1]} = \frac{1}{\sqrt{[2]\{A\}}}\left(\begin{array}{cc}
 \sqrt{\{A/q\}}  & \sqrt{\{Aq\}} \\ \\ \sqrt{\{Aq\}}  & -\sqrt{\{A/q\}}
\end{array}\right) = \frac{1}{\sqrt{[2][N]}}\left(\begin{array}{cc}
\sqrt{ [N-1]} & \sqrt{ [N+1]} \\ \\ \sqrt{ [N+1]} &- \sqrt{ [N-1]} \end{array}\right)
\label{S1}
\ee
It is symmetric, but, by essence, it is an illusion:
$S$ acts between {\it different} spaces, $R\otimes R$ and $R\otimes\bar R$,
thus, there is no actual sense in which it can be symmetric.

The eigenvalues of  $\mathfrak{h}^{[1]}$ are labeled by
$\bar X = \emptyset, \ {\rm Adj} \in [1]\otimes \overline{[1]}$:
\be
K^{[1]} = \left(\begin{array}{cc} K^{[1]}_{\emptyset} & 0 \\ 0 & K^{[1]}_{{\rm Adj}}\end{array}\right)
= \left(\begin{array}{cc} \frac{A}{\{q\}}\cdot(A^2 - q^2+1-q^{-2})  & 0 \\ 0 & -A\{q\}\end{array}\right)
\label{K1}
\ee
On the pretzel side, they are given by (\ref{KX}), i.e. are made from the truly symmetric matrix $\bar S$,
which we reconstruct from (\ref{S1}) with the help of (\ref{bSfromS})
with $T = -\frac{1}{qA} \left(\begin{array}{cc} 1& 0 \\ 0 & -q^2 \end{array}\right)$ and
$\bar T = \left(\begin{array}{cc} 1 & 0 \\ 0 & -A \end{array}\right)$:
\be
\bar S^{[1]} \stackrel{(\ref{bSfromS})}{=}\
\frac{\{q\}}{\{A\}} \left(\begin{array}{cc} 1 & \frac{\sqrt{\{Aq\}\{A/q\}}}{\{q\}} \\ \\
\frac{\sqrt{\{Aq\}\{A/q\}}}{\{q\}} & -1
\end{array}\right) =
\frac{1}{[N]} \left(\begin{array}{cc} 1 & \sqrt{[N-1][N+1]} \\ \\ \sqrt{[N-1][N+1]} & -1
\end{array}\right)
\label{bS1}
\ee
Substituting it  into (\ref{KX}),
together with $\lambda_\emptyset =1$,  $\lambda_{{\rm Adj}} = A$, $d_\emptyset = 1$, $d_{\rm Adj}
= \frac{ \{Aq\}\{A/q\}}{\{q\}^2} = \sqrt{[N-1][N+1]}$,
we reproduce (\ref{K1}):
\be
K_{\bar X} = d_R \frac{(\bar S \bar T^{2} \bar S)_{\emptyset \bar X}}{\sqrt{d_{\bar X}}} =
\left\{\begin{array}{ccll} \bar X=\emptyset:
&  d_{[1]}\Big(\bar S_{\emptyset\emptyset}^2 + \bar S_{\emptyset, {\rm Adj}}^2 A^{2}\Big)
& = \frac{1}{[N]}( 1 + A^{2}[N-1][N+1]) & = \frac{A^3 - A(q^2-1+q^{-2})}{\{q\}} \\ \\
\bar X={\rm Adj}:
&  \ \ d_{[1]}\frac{\bar S_{\emptyset\emptyset}
\bar S_{\emptyset, {\rm Adj}} }{\sqrt{d_{\rm Adj}}} (1- A^{2})
& \ = \ \ \ \  \frac{1 - A^{2}}{[N]} & \!\!\!\!\!\!\!\!\!\!\!\!\!\! = -A\{q\}
\end{array}
\right.
\nn
\ee
For $A=q^2$, i.e. for $sl_2$ the two matrices $S$ and $\bar S$ coincide,
while $T$ and $\bar T$ differ by a framing factor, which is actually essential,
because it does not drop out from (\ref{bSfromS}).
In this particular case it is equal to $qA=q^3$, and can be redistributed
in equal proportions between $T$ and two $\bar T$'s in (\ref{bSfromS}).

\subsection{Representation $R=[2]$}

This time $Y,Z\in [2]^{\otimes 2} = [22]\oplus[31]\oplus [4]$, dimensions are

{\footnotesize
\be
\frac{d_{[22]}}{d_{[2]}} = \frac{\{A\}\{A/q\}}{ \{q^2\}\{q^3\}} = \frac{[N][N-1]}{[2][3]}
\ \ \ \ \ \ \ \
\frac{d_{[31]}}{d_{[2]}}=\frac{\{Aq^2\}\{A/q\}}{ \{q\}\{q^4\}} = \frac{[N+2][N-1]}{[4]},
\ \ \ \ \ \ \ \
\frac{d_{[4]}}{d_{[2]}} = \frac{\{Aq^2\}\{Aq^3\}}{ \{q^3\}\{q^4\}} = \frac{[N+2][N+3]}{[3][4]}
\nn
\ee
}
and from (\ref{3str})
\be
\mathfrak{h}^{[2]}_{YZ} = \frac{d_{[2]}}{\sqrt{d_Yd_Z}}\cdot h^{[2]}_{YZ} =
\label{h2}
\ee
{\tiny
\be
=
\left(\begin{array}{c|ccc}
& [22] & [31] &[4] \\
\hline
&&&\\
\l[22]& \frac{A^2\Big(q^{10}A^4-q^5[2](q^8-q^2+1)A^2 + 1+ q^6[2][3](q^6-q^4+1)\{q\}\Big)}{q^7[2][3]\{q\}^2}
& \frac{A^3(q^6A^2-q^{10}+q^2-1)}{q^3\{q\}^2}\sqrt{\frac{\{Aq^2\}\{A\}}{[2][3][4]}}
& \frac{q^4A^4}{[3]\{q\}^2}\sqrt{\frac{\{Aq^3\}\{Aq^2\}\{A\}\{A/q\}}{[2][4]}} \\ &&&\\
\l[31] &  \frac{A^3(q^6A^2-q^{10}+q^2-1)}{q^3\{q\}^2}\sqrt{\frac{\{Aq^2\}\{A\}}{[2][3][4]}}
&\frac{A^2\Big(q^{12}A^4-q^7[2]\alpha_1A^2 +1+q^3\alpha_2\{q\}\Big)}
{q^7[4]\{q\}^2}
  & \frac{A^3(q^8A^2-q^8+q^4-1)}{q^2[4]\{q\}^2}\sqrt{\frac{\{Aq^3\}\{A/q\}}{[3]}} \\ &&&\\
\l[4]& \frac{q^4A^4}{[3]\{q\}^2}\sqrt{\frac{\{Aq^3\}\{Aq^2\}\{A\}\{A/q\}}{[2][4]}}
&  \frac{A^3(q^8A^2-q^8+q^4-1)}{q^2[4]\{q\}^2}\sqrt{\frac{\{Aq^3\}\{A/q\}}{[3]}}
& \frac{A^2\Big(q^{18}A^4-q^9[2](q^{10}-q^6+1)A^2 +1+q^8[2]\alpha_3\{q\}\Big)}{q^9[3][4]\{q\}^2}
\end{array}\right)
\nn
\ee}

{\footnotesize
$\alpha_1=(q^8-2q^6+4q^4-4q^2+2)$, $\alpha_2=(q^{12}-q^{10}+q^8+q^6-q^4+3q^2+1)$, $\alpha_3=(q^{10}-q^8+q^2+1)$.}

\bigskip

\noindent
It is diagonalized by the orthogonal matrix
\be
S_{[2]} = 
 \left(\begin{array}{ccc}
 \sqrt{{D_{-1}\over [3]D_1}}& \sqrt{[2]D_2D_{-1}\over [4]D_0D_1} & \sqrt{[2]D_2D_3\over [3][4]D_0D_1} \\
&&\\
{1\over\sqrt{[3]}} & (D_2-D_0)\sqrt{[2]\over [4]D_0D_2}
&  -[2]\sqrt{[2]D_{-1}D_3\over [3][4]D_0D_2}  \\
&&\\
\sqrt{D_3\over [3]D_1} &  -\sqrt{[2]D_0D_3\over [4]D_1D_2} &  \sqrt{[2]D_0D_{-1}\over [3][4]D_1D_2}
\end{array}\right)
\label{S2}
\ee
The eigenvalues of (\ref{h2}),
 $K_{\bar X}$ are labeled by $\bar X= \emptyset, {\rm Adj}=[2,1^{N-1}], [2,2,1^{N-2}]$ and equal to
\be
K^{[2]}_\emptyset
&=& \frac{A^2\Big(A^4q^{10}-A^2(q^{12}+q^{10}-q^8+q^4)+ q^{12}-q^{10}+2q^6-q^4-q^2+1\Big)}{q^5[2]\{q\}^2} \nn \\
K^{[2]}_{\rm Adj} &=& -q^{-2}A^2(A^2q^6-1+q^4-q^6) \nn \\
K^{[2]}_{[2,2,1^{N-2}]} &=& qA^2[2]\{q\}^2
\label{ev2}
\ee
Now one can construct from (\ref{S2}) by the rule (\ref{bSfromS}) with
\be
T^{[2]}=\frac{1}{q^4A^2}\!\!\left(\begin{array}{ccc} 1&&\\&-q^2\\&&q^6 \end{array}\right)\ \ \ \ \ \ \ \ 
\bar T^{[2]}=\left(\begin{array}{ccc} 1 && \\ & -A & \\ &&  q^2A^2  \end{array}\right)
\ee
the second exclusive matrix
\be
\bar S^{[2]}  \ \stackrel{(\ref{bSfromS})}{=}\
  \left(\begin{array}{ccc}
  \frac{[2]}{D_0D_1} & \frac{[2]}{D_0}\sqrt{\frac{D_{-1}}{D_1}} &\frac{\sqrt{D_{-1}D_3}}{D_1} \\ \\
\frac{[2]}{D_0}\sqrt{\frac{D_{-1}}{D_1}}
& \frac{D_0D_2-[2]^2}{D_0D_2}
& -\frac{[2]}{D_2}\sqrt{\frac{D_3}{D_1}} \\ \\
\frac{\sqrt{D_{-1}D_3}}{D_1} & -\frac{[2]}{D_2}\sqrt{\frac{D_3}{D_1}} & \frac{[2]}{D_1D_2}
\end{array}\right)
\ee

\subsection{Representation $R=[2,2]$\label{sS22}}

In this case formulas become much more tedious. The indices are now: $Y,Z\in [2,2]^{\otimes 2}=[4,4]\oplus [4,3,1]\oplus [4,2,2]\oplus [3,3,1,1]\oplus [3,2,2,1]\oplus [2,2,2,2]$ and

\be
\mathfrak{h}^{[22]}_{YZ} = \frac{\sqrt{d_Yd_Z}}{d_{[22]}}\cdot A^8\times
\ee

{\tiny
\be
\hspace{-1cm}\begin{array}{c||c|c|c|}
&[2222]&[3221]&[3311]\\
&&&\\
\hline \hline
&&&\\
\l[2222]& \beta_1
&  &   \\
&&&\\
\hline
&&&\\
\l[3221]& 1 - \frac{[2][3][4]t}{D_{-2}} + \frac{[3]^2[4]^2t^2}{D_{-1}D_{-3}} - \frac{[2][3]^2[4][5]t^3}{D_{-1}D_{-2}D_{-3}}
& \beta_2&
 \\
 &&&\\
\hline
&&&\\
\l[3311]& 1 - \frac{[2]^2[5]t}{D_{-2}} + \frac{[2][5][6](D_{-2}-D_0)t^2}{D_{-2}D_{-3}(D_{-1}-D_{1})} + \frac{[2][5][6]t^3}{D_{-3}D_{-2}(D_{-1}-D_{1})}
& 1 - \frac{[3]([5]D_{-2}+D_2)t}{D_2D_{-2}} + \frac{[3]^2[5]([4]D_0-2D_{1})t^2}{D_{-2}D_2D_{-3}} - \frac{[3]^2[5][6]t^3}{D_{-2}D_2D_{-3}}
 &\beta_3  \\
 &&&\\
\hline
&&&\\
 \l[422]& 1-\frac{[2]^2[3]t}{D_{-2}}+\frac{[2]^2[3]^2(D_0-D_{2})t^2}{D_{-2}D_{-1}(D_{-1}-D_{1})}
& 1 - \frac{([3]D_{-4}+[2]^3D_{1}+[3]D_{4})t}{D_2D_{-2}} + \frac{[3]^2([5]D_{-1}+D_{3})t^2}{D_{-2}D_{-1}D_2} - \frac{[2][3]^2[5]t^3}{D_{-2}D_{-1}D_2}& 1 - \frac{[3][5]D_0t}{D_{-2}D_2} + \frac{[3]^2[5]t^2}{D_{-2}D_2}
  \\
 &&&\\
\hline
&&&\\
\l[431]& 1 - \frac{[2][4]t}{D_{-2}}
& 1 - \frac{[3][4]^2D_0t}{[2]^2D_{-2}D_2} + \frac{[3]^2[4]^2t^2}{[2]^2D_{-2}D_2}&
  \\
 &&&\\
\hline
&&&\\
\l[44]& 1    & 1-\frac{[2][4]t}{D_{2}}&1 - \frac{[2]^2[3]t}{D_{2}}  + \frac{[2]^2[3]^2(D_{-2}-D_0)t^2}{D_{1}D_{2}(D_{-1}-D_{1}))}
 \\
&&&\\
\end{array}\nn
\ee
}
{\tiny
\be
\beta_1=1 - \frac{[2]^3[4]t}{D_{-2}} + \frac{[3][4]^2([2][5]+[2]^2D_{-2}  (D_{-1}-D_{1})+[3]D_{-1}(D_{-2}-D_0))t^2}{D_{-1}D_{-2}D_{-3}(D_{-1}-D_{1})} - 
\frac{[2][3][4]([4]D_{-4}+[3]^2([2][4]-1)(D_{-1}-D_{1}))t^3}{D_{-1}D_{-2}D_{-3}(D_{-1}-D_{1})} + \frac{[2]^2[3]^2[4]^2[5]t^4}{D_{-1}D_{-2}^2D_{-3}}\nn\\
\beta_2=1 - \frac{[4]^2([2]^2D_{2}+D_0)t}{[2]^2D_{-2}D_2} + \frac{[3]^2[4]^2([2]D_{-1}D_{-2}  +D_{2}D_0-[3]-[2]^2)t^2}{[2]^2D_{-2}D_{-1}D_2D_{-3}} -\frac{[3]^2[4]([5]D_{-3}+[3]D_{-1}+2D_{5})t^3}{D_{-2}D_{-1}D_2D_{-3}} + \frac{[2][3][4]^2[6]t^4}{D_{-2}D_{-1}D_2D_{-3}}\nn\\
\beta_3=1 - \frac{\Big([2]D_{5}+[7]D_{2}+[2]^3D_{-1}\Big)t}{D_{-2}D_2} + \frac{[3]\Delta_1t^2}{D_{-2}D_{1}D_2D_{-3}(D_{-1}-D_{1})} -
\frac{[2][3]\Delta_2t^3}{D_{-2}D_{1}D_2D_{-3}(D_{-1}-D_{1})} + \frac{[2]^3[3][5][6]t^4}{D_{-2}D_{1}D_2D_{-3}}\nn\\
\Delta_1=\{q\}^{-2}\left[A^{-3}\big(q^{-10}+q^{-8}+q^{-6}+4q^{-4}+7q^{-2}+5+5q^{2}+7q^{4}+5q^{6}+q^{8}\big)+A^{-1}\big(-q^{-10}-5q^{-8} -5q^{-6}-3q^{-4}-5q^{-2}-9-4q^{2}-q^{6}-3q^{8}-q^{10}\big)+\nn\right.\\ +
A\big(-q^{-10}-3q^{-8}-q^{-6}-4q^{-2}-9-5q^{2}-3q^{4}-5q^{6} -5q^{8}-q^{10}\big)+A^{3}\big(q^{-8}+5q^{-6}+7q^{-4}+5q^{-2}+5+7q^{2}+4q^{4}+q^{6}+q^{8}+q^{10}\big)\Big]\nn\\
\Delta_2=\{q\}^{-2}\left[A^{-2}\big(-q^{-12}-2q^{-10}-q^{-8}-q^{-6}-4q^{-4}-2q^{-2}+2+2q^{2}+q^{4}+4q^{6}+4q^{8}+2q^{10}\big)+ \big(-q^{-12}+2q^{-8}+q^{-6}-4q^{-4}-\nn\right.\\ -2q^{-2}-2q^{2}-4q^{4}+q^{6}+2q^{8}-q^{12}\big)+A^{2}\big(2q^{-10}+4q^{-8}+4q^{-6}+q^{-4} +2q^{-2}+2-2q^{2}-4q^{4}-q^{6}-q^{8}-2q^{10}-q^{12}\big)\Big]\nn
\ee
}

\noindent
Due to the symmetries $h_{YZ}=h_{ZY}$ and $h_{Y^{tr}Z^{tr}}(q)=h_{YZ}(q^{-1})$, it is sufficient to calculate only one quarter
of the table.

The six eigenvalues of $\mathfrak{h}^{[22]}_{YZ} = \frac{d_{[22]}}{\sqrt{d_Yd_Z}}\,h^{[22]}_{YZ}$ are associated with $\emptyset$, $[2,1^{N-2}]$, $[2,2,1^{N-4}]$, $[4,2^{N-2}]$, $[4,3,2^{N-4},1]$ and $[4,4,2^{N-4}]$:
{\footnotesize
\be
K_\emptyset^{[22]}&=&{ {A}^{4}\over [3][2]^2 \{q\}^4}
\left[ A^8-A^6q^{-6}\Big(1+2{q}^{2}-2{q}^{6}+2{q}^{10}+{q}^{12}\Big)+
A^4q^{-10}\Big(2{q}^{20}+{q}^{18}-2{q}^{14}-{q}^{16}+{q}^{12}+4{q}^{10}+{q}^{8}-2{q}^{6}-\right.\nn\\&& -{q}^{4}+{q}^{2}+2\Big) -A^2\Big({q}^{14}+{q}^{12}-2{q}^{10}-2{q}^{8}+3{q}^{6}+5{q}^{4}-{q}^{2}-6-{q}^{-2}+5{q}^{-4}+3{q}^{-6}-2{q}^{-8}-2{q}^{-10}+ {q}^{-12}+{q}^{-14}\Big)+\nn\\ &&+{q}^{16}-{q}^{14}-2{q}^{12}+2 {q}^{10}
+3{q}^{8}-{q}^{6}-4{q}^{4}+5-4{q}^{-4}-{q}^{-6}+3{q}^{-8}+2{q}^{-10}-2{q}^{-12}-{q}^{-14}+q^{-16} \Big]\nn
\\
K_{[2,1^{N-2}]}^{[22]}&=&{ {A}^{4}\over [3]\{q\}^2}\Big[A^6 -A^4\Big(q^6+q^4-1+q^{-4}+q^{-6}\Big)-A^2\Big(q^{10}-q^4+q^2+1+q^{-2}-q^{-4}+q^{-10}\Big) -\nn\\ &&-q^{12}+q^{10}+q^8-q^6-q^4-q^2+3-q^{-2}-q^{-4}-q^{-6}+q^{-8}+q^{-10}-q^{-12}\Big]\nn
\ee}
\vspace{-.7cm}
\be
K_{[2,2,1^{N-4}]}^{[22]}&=&q^{-8}A^4\Big(q^{10}A^4-q^5[2](q^8-q^6+1)A^2+(q^{16}-q^{14}-q^{12}+q^{10}+q^8-q^4+1)\Big) \nn\\
K_{[4,2^{N-2}]}^{[22]}&=&q^{-8}A^4\Big(q^{6}A^4-q^3[2](q^8-q^2+1)A^2+(q^{16} -q^{12} +q^8+q^6-q^4-q^2+1)\Big) \nn \\
K_{[4,3,2^{N-4},1]}^{[22]}&=&-[3]A^4(A^2-q^4+1-q^{-4})\{q\}^2 \nn \\
K_{[4,4,2^{N-4}]}^{[22]}&=&[2]^2[3]A^4\{q\}^4
\ee
The eigenvector, associated with $K_\emptyset^{[22]}$ is $\frac{\sqrt{d_Z}}{d_{[22]}}$. These eigenvectors are obtained from $\mathfrak{h}_{YZ}$ by rotating with the orthogonal Racah matrix

{\tiny
\be\hspace{-1cm}
S^{[2,2]} = \frac{1}{d_{[22]}}\left(\begin{array}{cccccc}
\sqrt{d_{[2222]}} & \sqrt{d_{[3221]}} & \sqrt{d_{[3311]}} & 63 & -62 & 61 \\ \\

-\sqrt{\frac{D_{-1}D_{-3}}{[5]}}\frac{D_1D_0D_{-1}}{[4][3]}  &  \sqrt{\frac{[2][3]D_{-1}D_{-3}}{[6]D_2D_{-2}}}\frac{d_{[22]}}{[4]D_0}(D_{-2}{-}D_4{-}D_2{-}D_0) & -\sqrt{\frac{[3][2]D_{1}D_{-3}}{[6][5]D_2D_{-2}}}\frac{D_1D_0D_{-1}}{[3][2]^3}(D_4{+}D_2{-}D_{-2}) & 53 & -52 & 51 \\ \\

\frac{1}{\sqrt{[5]}}\frac{D_1D_0^2D_{-1}}{[4][3][2]} & \sqrt{\frac{[3][2]}{[6]D_2D_{-2}}}\frac{D_1D_0^2D_{-1}}{[4][2]^2}(D_1-D_{-1}) & \sqrt{\frac{[3][2]D_1D_{-1}}{[6][5]D_2D_{-2}}}\frac{D_0^2}{[3][2]^3}(D_3D_1{+}D_{-2}^2{-}D_2D_{-2})  & -43 & 42 & -41 \\ \\

\sqrt{\frac{D_3D_1D_{-1}D_{-3}}{[5]}}\frac{D_0^2}{[4][2]} & -\sqrt{\frac{[3][2]D_3D_1D_{-1}D_{-3}}{[6]D_2D_{-2}}}\frac{D_0^2}{[4][2]^2}(D_1-D_{-1}) & \sqrt{\frac{[3][2]D_3D_{-3}}{[6][5]D_2D_{-2}}}\frac{D_1D_0^2D_{-1}}{[2]^3} & -33 & 32 & -31 \\ \\

\sqrt{\frac{D_3D_1}{[5]}}\frac{D_1D_0D_{-1}}{[4][3]} & \sqrt{\frac{[3][2]D_3D_1}{[6]D_2D_{-2}}}\frac{D_1D_0D_{-1}}{[4][3][2]^2}(D_2{-}[3]D_{-2}) & \sqrt{\frac{[2]D_3D_{-1}}{[6][5][3]D_2D_{-2}}}\frac{D_1D_0D_{-1}}{[2]^3}([2]D_{-3}{-}D_2) & 23 & -22 & 21 \\ \\

\sqrt{d_{[44]}} & -\sqrt{d_{[431]}} & \sqrt{d_{[422]}} & 13 & -12 & 11
\end{array}\right)\nn
\ee
}
The indices of this matrix are:
\be
\hbox{columns:}&\ \  (1) \mapsto [2,2,2,2],\ \   (2) \mapsto [3,2,2,1],\ \    (3) \mapsto [3,3,1,1],\ \   (4) \mapsto [4,2,2],\ \   (5) \mapsto [4,3,1],\ \   (6) \mapsto [4,4],\nn\\
\hbox{lines:}&\ \ (1) \mapsto \emptyset,\ \   (2) \mapsto [2,1^{N-2}],\ \    (3) \mapsto [2,2,1^{N-4}],\ \   (4) \mapsto [4,2^{N-2}],\ \   (5) \mapsto [4,3,2^{N-4},1],\ \   (6) \mapsto [4,4,2^{N-4}]\nn
\ee
It celebrates the symmetry $S_{ij}=\pm S_{7-i,7-j}$.

The corresponding dimensions are:
{\footnotesize
\be
d_{[2222]} = \frac{D_{-3}D_{-2}^2D_{-1}^2D_0^2D_1}{[5][4]^2[3]^2[2]^2}, \ \ \ \ \ \ 
d_{[3221]} = \frac{D_{-3}D_{-2}D_{-1}^2D_0^2D_1D_2}{[6][4]^2[3][2]}, \ \ \ \ \ \
d_{[3311]} = \frac{D_{-3}D_{-2}D_{-1}D_0^2D_1^2D_2}{[6][5][3][2]^3},\nn\\
d_{[422]}  = \frac{D_{-2}D_{-1}^2D_0^2D_1D_2D_3}{[6][5][3][2]^3}, \ \ \ \ \ \
d_{[431]}  = \frac{D_{-2}D_{-1}D_0^2D_1^2D_2D_3}{[6][4]^2[3][2]}, \ \ \ \ \ \
d_{[44]}   = \frac{D_{-1}D_0^2D_1^2D_2^2D_3}{[5][4]^2[3]^2[2]^2}
\ee}
This matrix $S$ is made entirely of quantum numbers. Now using the $T$-matrices
\be
{T}^{[2,2]} = \hbox{diag}\Big(A^4q^8,\ -A^4q^4,\ A^4q^2,\ \frac{A^4}{q^2},\ -\frac{A^4}{q^4},\ \frac{A^4}{q^8}\Big)\nn\\
{\bar T}^{[2,2]} =\hbox{diag}\Big(1,\ -\frac{1}{A},\ \frac{q^2}{A^2},\ \frac{1}{A^2q^2},\ -\frac{1}{A^3},\ \frac{1}{A^4}\Big)
\ee
and (\ref{bSfromS}), one obtains the second exclusive matrix $\bar S^{[2,2]}$:

{\footnotesize
\be
\bar S^{[2,2]} = \frac{1}{d_{[2,2]}}\left(\begin{array}{cccccc}
\sqrt{\bar d_1} & -\sqrt{\bar d_2} & \sqrt{\bar d_3} & -\sqrt{\bar d_4} & \sqrt{\bar d_5} & \sqrt{\bar d_6} \\ \\

12 & \frac{D_1D_{-1}}{[2]^2D_2D_{-2}}\gamma_1 & -\frac{\sqrt{D_{-3}D_{-1}}D_1D_0}{[2]^2D_2D_{-2}}\gamma_2 & \frac{\sqrt{D_{3}D_{1}}D_0D_{-1}}{[2]^2D_2D_{-2}}\gamma_3 & -\frac{\sqrt{D_{3}D_{1}D_{-1}D_{-3}}D_1D_{-1}}{[2]^2[3]D_2D_{-2}}\gamma_4 & 15 \\ \\

13 & 23 & \frac{D_0^2}{[2]^2[3]D_2D_{-2}}\gamma_5 & \frac{\sqrt{D_{3}D_{1}D_{-1}D_{-3}}D_0^2[3]}{[2]^2D_2D_{-2}} & -24 & 14 \\ \\

14 & 24 & 34 & 33 & -23 & -13 \\ \\

15 & 25 & 35 & 45 & 22 & 12 \\ \\

16 & 26 & 36 & 46 & 56 & 11
\end{array}\right)
\ee

\bigskip

\noindent
with $\gamma_1=[3]D_2D_{-2} - [2]^2, \ \gamma_2=D_3D_{-2} - [2], \ \gamma_3=D_2D_{-3} - [2], \ \gamma_4=D_3D_{-3} - 2[3] - 1, \ \gamma_5=D_3D_2D_{-2}D_{-3} - D_{3}D_{-3} + [2]^2$.
}
Here $\bar d_i$ are quantum dimensions of the corresponding representations:
{\footnotesize
\be
\bar d_1=1, \ \ \ \ \ \ \ 
\bar d_2=D_1D_{-1}, \ \ \ \ \ \ \ 
\bar d_3=\frac{D_{-3}D_{0}^2D_{1}}{[2]^2}, \ \ \ \ \ \ \ 
\bar d_4=\frac{D_{-1}D_{0}^2D_{3}}{[2]^2}, \ \ \ \ \ \ \ 
\bar d_5=\frac{D_{-3}D_{-1}^2D_{1}^2D_3}{[3]^2}, \ \ \ \ \ \ \ 
\bar d_6=\prod_{i=-3}^{3}\frac{D_{-2}D_2}{[3]^2[2]^4D_0}
\ee}
The matrix has two symmetries: $\bar S_{i,j} = \bar S_{j,i}$ and $\bar S_{i,j} = \pm\bar S_{7-j,7-i}$, the signs are given explicitly if needed.

\section{Examples of $[2,2]$-colored HOMFLY
for knots, which are arborescent but not 3-strand}

Using the manifest expressions for the exclusive Racah matrices $S^{[2,2]}$ and $\bar S^{[2,2]}$ in sec.\ref{sS22}, one can evaluate the HOMFLY polynomials of all arborescent knots in representation $[2,2]$. The results for simplest knots have been found in \cite{Naw,modRT2}, those for more complicated knots can be found in \cite{knotebook}, here, as an illustration, we write down the answers for three knots that are arborescent and can not be presented by a 3-strand braid (when an equivalent evaluation by methods of \cite{MMMS22} is available).

\paragraph{Knot $6_1$, braid index 4:}

{\footnotesize
$H_{[2,2]}^{6_1}=\Big( {A}^{24}{q}^{20}+ \Big( -{q}^{26}-2\,{q}^{24}+2\,{q}^{20}-2\,{q}^{16}-{q}^{14} \Big) {A}^{22}+ \Big( 2\,{q}^{30}+{q}^{28}-2\,{q}^{26}-4\,{q}^{24}+2\,{q}^{22}+8\,{q}^{20}+2\,{q}^{18}-4\,{q}^{16}-2\,{q}^{14}+{q}^{12}+2\,{q}^{10} \Big) {A}^{20}+ \Big( -{q}^{34}+4\,{q}^{30}+2\,{q}^{28}-6\,{q}^{26}-8\,{q}^{24}+3\,{q}^{22}+12\,{q}^{20}+3\,{q}^{18}-8\,{q}^{16}-6\,{q}^{14}+2\,{q}^{12}+4\,{q}^{10}-{q}^{6} \Big) {A}^{18}+ \Big( -2\,{q}^{34}-{q}^{32}+6\,{q}^{30}+3\,{q}^{28}-10\,{q}^{26}-13\,{q}^{24}+3\,{q}^{22}+17\,{q}^{20}+3\,{q}^{18}-13\,{q}^{16}-10\,{q}^{14}+3\,{q}^{12}+6\,{q}^{10}-{q}^{8}-2\,{q}^{6} \Big) {A}^{16}+ \Big( -2\,{q}^{34}+9\,{q}^{30}+5\,{q}^{28}-11\,{q}^{26}-14\,{q}^{24}+7\,{q}^{22}+24\,{q}^{20}+7\,{q}^{18}-14\,{q}^{16}-11\,{q}^{14}+5\,{q}^{12}+9\,{q}^{10}-2\,{q}^{6} \Big) {A}^{14}+ \Big( {q}^{40}-{q}^{36}-3\,{q}^{34}+{q}^{32}+11\,{q}^{30}+4\,{q}^{28}-15\,{q}^{26}-18\,{q}^{24}+7\,{q}^{22}+28\,{q}^{20}+7\,{q}^{18}-18\,{q}^{16}-15\,{q}^{14}+4\,{q}^{12}+11\,{q}^{10}+{q}^{8}-3\,{q}^{6}-{q}^{4}+1 \Big) {A}^{12}+ \Big( -{q}^{36}-4\,{q}^{34}+11\,{q}^{30}+5\,{q}^{28}-18\,{q}^{26}-23\,{q}^{24}+8\,{q}^{22}+32\,{q}^{20}+8\,{q}^{18}-23\,{q}^{16}-18\,{q}^{14}+5\,{q}^{12}+11\,{q}^{10}-4\,{q}^{6}-{q}^{4} \Big) {A}^{10}+ \Big( -2\,{q}^{34}+11\,{q}^{30}+5\,{q}^{28}-15\,{q}^{26}-20\,{q}^{24}+8\,{q}^{22}+32\,{q}^{20}+8\,{q}^{18}-20\,{q}^{16}-15\,{q}^{14}+5\,{q}^{12}+11\,{q}^{10}-2\,{q}^{6} \Big) {A}^{8}+ \Big( -{q}^{34}+{q}^{32}+6\,{q}^{30}+2\,{q}^{28}-10\,{q}^{26}-12\,{q}^{24}+6\,{q}^{22}+20\,{q}^{20}+6\,{q}^{18}-12\,{q}^{16}-10\,{q}^{14}+2\,{q}^{12}+6\,{q}^{10}+{q}^{8}-{q}^{6} \Big) {A}^{6}+ \Big( 2\,{q}^{30}-5\,{q}^{26}-6\,{q}^{24}+2\,{q}^{22}+10\,{q}^{20}+2\,{q}^{18}-6\,{q}^{16}-5\,{q}^{14}+2\,{q}^{10} \Big) {A}^{4}+ \Big( -{q}^{26}-2\,{q}^{24}+{q}^{22}+4\,{q}^{20}+{q}^{18}-2\,{q}^{16}-{q}^{14} \Big) {A}^{2}+{q}^{20} \Big){{q}^{-20}{A}^{-8}}
$
}

\paragraph{Knot $9_{46}$, braid index 4:}

{\footnotesize
$H_{[2,2]}^{9_{46}}=\Big( {A}^{24}{q}^{24}+ \Big( -{q}^{30}-2\,{q}^{28}+2\,{q}^{24}-2\,{q}^{20}-{q}^{18} \Big) {A}^{22}+ \Big( 2\,{q}^{34}+{q}^{32}-2\,{q}^{30}-4\,{q}^{28}+{q}^{26}+6\,{q}^{24}+{q}^{22}-4\,{q}^{20}-2\,{q}^{18}+{q}^{16}+2\,{q}^{14} \Big) {A}^{20}+ \Big( -{q}^{38}+4\,{q}^{34}+3\,{q}^{32}-3\,{q}^{30}-5\,{q}^{28}+4\,{q}^{26}+12\,{q}^{24}+4\,{q}^{22}-5\,{q}^{20}-3\,{q}^{18}+3\,{q}^{16}+4\,{q}^{14}-{q}^{10} \Big) {A}^{18}+ \Big( -2\,{q}^{38}-2\,{q}^{36}+3\,{q}^{34}+{q}^{32}-8\,{q}^{30}-11\,{q}^{28}+9\,{q}^{24}-11\,{q}^{20}-8\,{q}^{18}+{q}^{16}+3\,{q}^{14}-2\,{q}^{12}-2\,{q}^{10} \Big) {A}^{16}+ \Big( {q}^{40}-{q}^{38}-2\,{q}^{36}+5\,{q}^{34}+4\,{q}^{32}-7\,{q}^{30}-12\,{q}^{28}+{q}^{26}+14\,{q}^{24}+{q}^{22}-12\,{q}^{20}-7\,{q}^{18}+4\,{q}^{16}+5\,{q}^{14}-2\,{q}^{12}-{q}^{10}+{q}^{8} \Big) {A}^{14}+ \Big( {q}^{40}-{q}^{38}+9\,{q}^{34}+11\,{q}^{32}-2\,{q}^{30}-6\,{q}^{28}+9\,{q}^{26}+24\,{q}^{24}+9\,{q}^{22}-6\,{q}^{20}-2\,{q}^{18}+11\,{q}^{16}+9\,{q}^{14}-{q}^{10}+{q}^{8} \Big) {A}^{12}+ \Big( {q}^{42}+{q}^{40}-4\,{q}^{38}-6\,{q}^{36}+4\,{q}^{34}+5\,{q}^{32}-9\,{q}^{30}-17\,{q}^{28}-3\,{q}^{26}+12\,{q}^{24}-3\,{q}^{22}-17\,{q}^{20}-9\,{q}^{18}+5\,{q}^{16}+4\,{q}^{14}-6\,{q}^{12}-4\,{q}^{10}+{q}^{8}+{q}^{6} \Big) {A}^{10}+ \Big( {q}^{48}-{q}^{46}-2\,{q}^{44}+{q}^{42}+3\,{q}^{40}-3\,{q}^{38}-9\,{q}^{36}+{q}^{34}+5\,{q}^{32}-6\,{q}^{30}-15\,{q}^{28}-3\,{q}^{26}+9\,{q}^{24}-3\,{q}^{22}-15\,{q}^{20}-6\,{q}^{18}+5\,{q}^{16}+{q}^{14}-9\,{q}^{12}-3\,{q}^{10}+3\,{q}^{8}+{q}^{6}-2\,{q}^{4}-{q}^{2}+1 \Big) {A}^{8}+ \Big( -{q}^{46}-{q}^{44}+{q}^{42}+4\,{q}^{40}-3\,{q}^{36}+5\,{q}^{34}+10\,{q}^{32}+5\,{q}^{30}-{q}^{28}+8\,{q}^{26}+18\,{q}^{24}+8\,{q}^{22}-{q}^{20}+5\,{q}^{18}+10\,{q}^{16}+5\,{q}^{14}-3\,{q}^{12}+4\,{q}^{8}+{q}^{6}-{q}^{4}-{q}^{2} \Big) {A}^{6}+ \Big( 2\,{q}^{42}+2\,{q}^{40}-3\,{q}^{36}+{q}^{34}+3\,{q}^{32}-3\,{q}^{30}-9\,{q}^{28}-{q}^{26}+8\,{q}^{24}-{q}^{22}-9\,{q}^{20}-3\,{q}^{18}+3\,{q}^{16}+{q}^{14}-3\,{q}^{12}+2\,{q}^{8}+2\,{q}^{6} \Big) {A}^{4}+ \Big( -{q}^{38}-3\,{q}^{36}-{q}^{34}-4\,{q}^{30}-8\,{q}^{28}-2\,{q}^{26}+6\,{q}^{24}-2\,{q}^{22}-8\,{q}^{20}-4\,{q}^{18}-{q}^{14}-3\,{q}^{12}-{q}^{10} \Big) {A}^{2}+{q}^{32}+{q}^{30}+{q}^{28}+2\,{q}^{26}+6\,{q}^{24}+2\,{q}^{22}+{q}^{20}+{q}^{18}+{q}^{16} \Big){q}^{-24}
$
}

\paragraph{Knot $10_{137}$, braid index 5:}

{\footnotesize
$H_{[2,2]}^{10_{137}}=\Big( {q}^{32}{A}^{32}-{q}^{26} \Big( 2\,{q}^{8}-3\,{q}^{4}+2 \Big)  \Big( {q}^{2}+1 \Big) ^{2}{A}^{30}+{q}^{18} \Big( {q}^{28}+{q}^{26}+5\,{q}^{24}+3\,{q}^{22}-8\,{q}^{20}-12\,{q}^{18}+6\,{q}^{16}+22\,{q}^{14}+6\,{q}^{12}-12\,{q}^{10}-8\,{q}^{8}+3\,{q}^{6}+5\,{q}^{4}+{q}^{2}+1 \Big) {A}^{28}-{q}^{12} \Big( {q}^{36}+2\,{q}^{30}-3\,{q}^{28}-5\,{q}^{26}+10\,{q}^{24}+7\,{q}^{22}+5\,{q}^{20}-25\,{q}^{18}+5\,{q}^{16}+7\,{q}^{14}+10\,{q}^{12}-5\,{q}^{10}-3\,{q}^{8}+2\,{q}^{6}+1 \Big)  \Big( {q}^{2}+1 \Big) ^{2}{A}^{26}+{q}^{8} \Big( 2\,{q}^{48}+{q}^{46}-{q}^{44}+7\,{q}^{40}-2\,{q}^{36}+24\,{q}^{34}+21\,{q}^{32}-28\,{q}^{30}-46\,{q}^{28}+21\,{q}^{26}+79\,{q}^{24}+21\,{q}^{22}-46\,{q}^{20}-28\,{q}^{18}+21\,{q}^{16}+24\,{q}^{14}-2\,{q}^{12}+7\,{q}^{8}-{q}^{4}+{q}^{2}+2 \Big) {A}^{24}-{q}^{4} \Big( {q}^{52}-{q}^{50}-{q}^{48}+6\,{q}^{46}-2\,{q}^{44}-6\,{q}^{40}+27\,{q}^{38}-24\,{q}^{36}-2\,{q}^{34}+52\,{q}^{30}-16\,{q}^{28}-33\,{q}^{26}-16\,{q}^{24}+52\,{q}^{22}-2\,{q}^{18}-24\,{q}^{16}+27\,{q}^{14}-6\,{q}^{12}-2\,{q}^{8}+6\,{q}^{6}-{q}^{4}-{q}^{2}+1 \Big)  \Big( {q}^{2}+1 \Big) ^{2}{A}^{22}+{q}^{2} \Big( {q}^{60}-{q}^{58}+3\,{q}^{56}+6\,{q}^{54}-3\,{q}^{52}-5\,{q}^{50}+15\,{q}^{48}+25\,{q}^{46}-11\,{q}^{44}-22\,{q}^{42}+54\,{q}^{40}+66\,{q}^{38}-38\,{q}^{36}-90\,{q}^{34}+36\,{q}^{32}+148\,{q}^{30}+36\,{q}^{28}-90\,{q}^{26}-38\,{q}^{24}+66\,{q}^{22}+54\,{q}^{20}-22\,{q}^{18}-11\,{q}^{16}+25\,{q}^{14}+15\,{q}^{12}-5\,{q}^{10}-3\,{q}^{8}+6\,{q}^{6}+3\,{q}^{4}-{q}^{2}+1 \Big) {A}^{20}-{q}^{2} \Big( {q}^{56}+2\,{q}^{54}-9\,{q}^{52}+13\,{q}^{50}+4\,{q}^{46}-24\,{q}^{44}+27\,{q}^{42}+22\,{q}^{40}-9\,{q}^{38}-53\,{q}^{36}+45\,{q}^{34}+46\,{q}^{32}+25\,{q}^{30}-105\,{q}^{28}+25\,{q}^{26}+46\,{q}^{24}+45\,{q}^{22}-53\,{q}^{20}-9\,{q}^{18}+22\,{q}^{16}+27\,{q}^{14}-24\,{q}^{12}+4\,{q}^{10}+13\,{q}^{6}-9\,{q}^{4}+2\,{q}^{2}+1 \Big)  \Big( {q}^{2}+1 \Big) ^{2}{A}^{18}+ \Big( {q}^{64}-{q}^{62}-{q}^{60}+10\,{q}^{58}+7\,{q}^{56}-18\,{q}^{54}-2\,{q}^{52}+48\,{q}^{50}+39\,{q}^{48}-55\,{q}^{46}-51\,{q}^{44}+113\,{q}^{42}+143\,{q}^{40}-68\,{q}^{38}-172\,{q}^{36}+58\,{q}^{34}+256\,{q}^{32}+58\,{q}^{30}-172\,{q}^{28}-68\,{q}^{26}+143\,{q}^{24}+113\,{q}^{22}-51\,{q}^{20}-55\,{q}^{18}+39\,{q}^{16}+48\,{q}^{14}-2\,{q}^{12}-18\,{q}^{10}+7\,{q}^{8}+10\,{q}^{6}-{q}^{4}-{q}^{2}+1 \Big) {A}^{16}-{q}^{2} \Big( 2\,{q}^{56}-3\,{q}^{54}-2\,{q}^{52}+11\,{q}^{50}+7\,{q}^{48}-20\,{q}^{46}-6\,{q}^{44}+24\,{q}^{42}+56\,{q}^{40}-56\,{q}^{38}-44\,{q}^{36}+31\,{q}^{34}+111\,{q}^{32}-13\,{q}^{30}-103\,{q}^{28}-13\,{q}^{26}+111\,{q}^{24}+31\,{q}^{22}-44\,{q}^{20}-56\,{q}^{18}+56\,{q}^{16}+24\,{q}^{14}-6\,{q}^{12}-20\,{q}^{10}+7\,{q}^{8}+11\,{q}^{6}-2\,{q}^{4}-3\,{q}^{2}+2 \Big)  \Big( {q}^{2}+1 \Big) ^{2}{A}^{14}+{q}^{4} \Big( {q}^{56}+6\,{q}^{54}-2\,{q}^{52}-12\,{q}^{50}+11\,{q}^{48}+45\,{q}^{46}+16\,{q}^{44}-66\,{q}^{42}-34\,{q}^{40}+132\,{q}^{38}+134\,{q}^{36}-86\,{q}^{34}-175\,{q}^{32}+67\,{q}^{30}+260\,{q}^{28}+67\,{q}^{26}-175\,{q}^{24}-86\,{q}^{22}+134\,{q}^{20}+132\,{q}^{18}-34\,{q}^{16}-66\,{q}^{14}+16\,{q}^{12}+45\,{q}^{10}+11\,{q}^{8}-12\,{q}^{6}-2\,{q}^{4}+6\,{q}^{2}+1 \Big) {A}^{12}-{q}^{8} \Big( 4\,{q}^{44}-7\,{q}^{40}-3\,{q}^{38}+21\,{q}^{36}+24\,{q}^{34}-33\,{q}^{32}-35\,{q}^{30}+38\,{q}^{28}+64\,{q}^{26}-82\,{q}^{22}+64\,{q}^{18}+38\,{q}^{16}-35\,{q}^{14}-33\,{q}^{12}+24\,{q}^{10}+21\,{q}^{8}-3\,{q}^{6}-7\,{q}^{4}+4 \Big)  \Big( {q}^{2}+1 \Big) ^{2}{A}^{10} \Big)  A^{-24}{q}^{-32}
$
}

\section{Integrable structure of $\mathfrak{h}$: $t$-decomposition}

\subsection{Fundamental representation}

As already observed in \cite{MMMS31}, the evolution matrices $h$ have a peculiar structure
of a polynomial in powers of $\{q\}$, somewhat reminiscent of the differential expansion
\cite{DGR,IMMMfe,evo,GGS,diffarth,Kondef},
which we are now going to reveal and exploit.
Namely,
\be
\mathfrak{h}^{[1]}_{YZ} = \frac{\sqrt{d_Yd_Z}}{d_{[1]}}\cdot A^2\cdot
\left(\begin{array}{cc} 1 -\frac{[2]\cdot t}{[N-1]} & 1 \\ 1 & 1-\frac{[2]\cdot t}{[N+1]}\end{array}\right)
\label{tH1t}
\ee
with $t=A^{-1}\{q\}$.
At $t=0$ the two normalized eigenvectors of
$\mathfrak{h}^{[1]}_{YZ}(t=0) \sim \sqrt{d_Yd_Z}$ with $\sum_Y d_Y = d_{[1]}$
form a symmetric orthogonal matrix
\be
\left.S^{[1]}\right|_{t=0} =
\left(\begin{array}{cc} \frac{\sqrt{d_{[11]}}}{d_{[1]}} &  \frac{\sqrt{d_{[2]}}}{d_{[1]}} \\
\frac{\sqrt{d_{[2]}}}{d_{[1]}} & - \frac{\sqrt{d_{[11]}}}{d_{[1]}}\end{array}\right)=
\left(\begin{array}{cc} \sqrt{\frac{[N-1]}{[2][N]}} &  \sqrt{\frac{[N+1]}{[2][N]}} \\
\sqrt{\frac{[N+1]}{[2][N]}} & - \sqrt{\frac{[N-1]}{[2][N]}}\end{array}\right)
\ee
which  surprisingly coincides with exact answer (\ref{S1}) true for all $t$.
In other words, the $t$-independent matrix $S^{[1]}$ from (\ref{S1}) diagonalizes $\mathfrak{h}^{[1]}$
from (\ref{tH1t}) at any $t$, and only the eigenvalues are $t$-dependent:
\be
K^{[1]} = \left(\begin{array}{cc} K^{[1]}_{\emptyset} & 0 \\ 0 & K^{[1]}_{{\rm Adj}}\end{array}\right)
= \left(\begin{array}{cc} \frac{A}{\{q\}}\cdot(A^2 - t\{q\}A -1)  & 0 \\ 0 & -A^2t\end{array}\right)
\ee
The resolution of the mystery is simple: the $t$-linear term in (\ref{tH1t})
is actually a unit matrix, therefore it leaves eigenvectors intact, but shifts the eigenvalues:
\be
\boxed{
\mathfrak{h}^{[1]}_{YZ} =A^2\cdot \left( \frac{\sqrt{d_Yd_Z}}{d_{[1]}}  -  t\cdot   \delta_{YZ}\right)
}
\label{tH1can}
\ee
One normalized eigenvector is obviously $\mu_Z = (v_1,v_2)=\frac{\sqrt{d_Z}}{d_{[1]}}$
with the eigenvalue $ A^2(d_{[1]}-t)$,
and the other one is its orthogonal complement $\mu^\bot_Z = (\mu_2,-\mu_1)$ with the eigenvalue $-tA^2$.
The matrix $S^{[1]}$ is made from these normalized eigenvectors in the usual way:
\be
S^{[1]} = \left(\begin{array}{cc} \mu_1 & \mu_2 \\ \mu_2 & -\mu_1 \end{array}\right)
\ee
The same structure preserves for more complicated representations.

\subsection{First symmetric representation}

Similarly to (\ref{tH1t}),
{\footnotesize
\be
\mathfrak{h}^{[2]}_{YZ} = \frac{\sqrt{d_Yd_Z}}{d_{[2]}}\cdot q^4 A^4\cdot
\left(\begin{array}{c|ccc}
& [22] & [31] &[4] \\
\hline
&&&\\
\l[22]& 1- \frac{q^{-2}[2][3]\cdot t}{[N-1]}+ \frac{q^{-3}[2]^2[3]\cdot t^2}{[N][N-1]}
&1 - \frac{q^{-2}[4]\cdot t}{[N-1]} & 1 \\ &&&\\
\l[31] & 1 - \frac{q^{-2}[4]\cdot t}{[N-1]}
& 1 - V\cdot\frac{q^{-10}[2]\cdot t}{[3]}    +v\cdot \frac{q^{-8}[2]\cdot t^2}{[N-1][N+2]}
  & 1     -\frac{q^{-2}[2]^2\cdot t}{[N+2]}\\ &&&\\
\l[4]& 1 & 1  -\frac{q^{-2}[2]^2\cdot t}{[N+2]}
& 1- \frac{q^{-2}[2]^2[3]\cdot t}{[N+2]}+ \frac{q^{-3}[2][3][4]\cdot t^2}{[N+2][N+3]}
\end{array}\right)
\nn
\ee
}
where
\be
V = \frac{q^3(q^{10}+q^6+q^4-q^2-v(q^2-1)}{[N+2]}+
\frac{q^{12}-q^{10}+2q^8+q^4+q^2+v(q^2-1)}{[N-1]}
\ee
with arbitrary $v$.
Having our experience with $R=[1]$, it is natural to observe that the
two out of three $t^2$-items in $\mathfrak{h}^{[2]}$ are equal.
Then, one can wish to make the entire $t^2$ contribution  proportional to the unit matrix,
for this one should put $v=q^5[4]$, and $\mathfrak{h}^{[2]}$ becomes
\be
\boxed{
\mathfrak{h}^{[2]}_{YZ} = q^4 A^4\cdot\left(\frac{\sqrt{d_Yd_Z}}{d_{[2]}} - t \cdot q^{-2}\cdot  \eta_{YZ}
+   t^2\cdot\frac{[2]}{q^3}\cdot\delta_{YZ}\right)
}
\label{tH2t}
\ee

\noindent
with the new matrix
\be
\eta =
\frac{\sqrt{d_Yd_Z}}{d_{[2]}}
\left(\begin{array}{ccc} \frac{ [2][3] }{[N-1]}
&  \frac{ [4] }{[N-1]} & 0 \\
 \frac{ [4] }{[N-1]}
&    \frac{[2]^2}{[3][N+2]} + \frac{[4]^2}{[2][3][N-1]}
  &   \frac{ [2]^2 }{[N+2]}\\
0 &     \frac{ [2]^2 }{[N+2]}
&     \frac{ [2]^2[3] }{[N+2]}
\end{array}\right) =
\left(\begin{array}{ccc} [N]
&   \sqrt{\frac{ [4][N][N+2] }{[2][3]}}  & 0 \\ \\
  \sqrt{\frac{ [4][N][N+2] }{[2][3]}}
&  \   \frac{[2]^2[N-1]}{[3][4]} + \frac{[4][N+2]}{[2][3]} \
  &    \frac{ [2]^2 }{[4]}\sqrt{\frac{[N+3][N-1]}{[3]}} \\ \\
0 &   \frac{ [2]^2 }{[4]}\sqrt{\frac{[N+3][N-1]}{[3]}}
& \frac{ [2]^2[N+3] }{[4]}
\end{array}\right)
\nn
\ee
As one could anticipate,
the eigenvector $\mu_Z = \frac{\sqrt{d_Z}}{d_{[2]}}$
of the unperturbed matrix $\mathfrak{h}_{YZ}(t=0)  \sim \sqrt{d_Yd_Z}$
(with which $\eta$ commutes)
remains to be the eigenvector of $\eta$ with eigenvalue $[2][N+1]$.
However, in the orthogonal space, $\eta$ specifies a preferred direction:
$\det(\eta)=0$ and the normalized zero mode $\vec \nu$ is
\be
\nu_1 = \frac{1}{\sqrt{3}}, \ \
\nu_2={D_3+D_{-1}\over [2]}\sqrt{\frac{[2]}{[4][N][N+2]}},\ \
\nu_3 = -[2]\sqrt{\frac{[2][N+3][N-1]}{[3][4][N][N+2]}} \nn \\
\ee
The remaining normalized eigenvector $\vec\rho$ is
\be
\rho_1 = \sqrt{\frac{[N+3]}{[3][N+1]}}, \ \
\rho_2 = -\sqrt{\frac{[2][N][N+3]}{[4][N+1][N+2]}}, \ \
\rho_3 = \sqrt{\frac{[2][N][N-1]}{[3][4][N+1][N+2]}}
\ee
its $\eta$-eigenvalue is $[N+2]$.
Thus, the three eigenvalues of $\mathfrak{h}^{[2]}$ are
\be
q^4A^4\left(d_{[2]} - \frac{t}{q^2} [2][N+1] + \frac{[2]t^2}{q^3}\right), \ \ \ \ \ \ \
q^4A^4\left( - \frac{t}{q^2} [N+2] + \frac{[2]t^2}{q^3}\right), \ \ \ \ \ \ \
q^4A^4\cdot \frac{[2]t^2}{q^3}
\label{ev2rev}
\ee
what reproduces (\ref{ev2}).

Now, one can immediately reproduce the Racah matrix, since the three columns of $S$ are made from the three vectors $\vec\mu,\vec\nu,\vec\rho$ which form an orthonormal basis. Indeed,
\be
S^{[2]}=\left(\begin{array}{ccc}
\mu_1&\mu_2&\mu_3\\
\nu_1&\nu_2&\nu_3\\
\rho_1&\rho_2&\rho_3
\end{array}\right)
\ee
is exactly equal to (\ref{S2}).
The original evolution matrix is expressed through these vectors in the following way:
\be
\mathfrak{h}^{[2]}_{YZ} = q^4 A^4\cdot\left(\frac{\sqrt{d_Yd_Z}}{d_{[2]}} - t \cdot q^{-2}\cdot  \eta_{YZ}
+   t^2\cdot\frac{[2]}{q^3}\cdot\delta_{YZ}\right)
= \nn \\
= q^4A^4\left\{d_{[2]}\cdot \mu_Y\mu_Z - t \cdot \frac{1}{q^2} \cdot
\Big( \underbrace{[2][N+1]\cdot   \mu_Y\mu_Z +  [N+2]\cdot  \nu_Y\nu_Z}_{\eta_{YZ}} \Big)
+   t^2\cdot\frac{[2]}{q^3}\cdot
\Big(\underbrace{\mu_Y\mu_Z + \nu_Y\nu_Z+\rho_Y\rho_Z}_{\delta_{YZ}}\Big)\right\}
\label{tHt2deco}
\ee
The rank two matrix $\eta$ is distinguished by being tri-diagonal
(note that one-diagonal is the
rank-three unit matrix, while the rank-one non-perturbed matrix has no zeroes
in the original basis).
Both commutativity and orthogonality properties are obvious from (\ref{tHt2deco}).

\subsection{General structure of $t\eta$-decomposition and hidden integrability\label{alg}}

Now we are ready to conjecture the general structure behind the problem.
\begin{itemize}
\item The evolution matrix for the family in the intersection $Pretzel \ \cap \ 3-strand$ is
\be
\mathfrak{h}^R_{YZ} = q^{4\varkappa_R} A^{2|R|} \left(\frac{\sqrt{d_Yd_Z}}{d_R} + \sum_{k=1}^{|R|-1}
(-t)^k \cdot \eta^{(k)}_{YZ} + (-t)^{|R|}\delta_{YZ}\right)
\ee
with $t= A^{-1}\{q\}$.
\item All matrices $\eta^{(k)}$ with $k=0,\ldots,|R|$ commute and have common $t$-independent
eigenvectors, which, being normalized form the orthogonal matrix $S$.
\item One of these normalized eigenvectors is always $\mu^{(0)}_Z = \frac{\sqrt{d_Z}}{d_R}$,
all others lie in the orthogonal space and they are graded by the condition that the eigenvalues are
of different non-vanishing orders in $t$.
\item The matrix $\eta^{(k)}$ has rank $k+1$ and at the same time
contains no more than $|R|-k+2$ non-zero sub-diagonals.
For symmetric representations, $R=[r]$ this estimate is exact.
\item $\eta^{(k)}$ is made out of $k+1$ unit vectors, which
are the first $k+1$ lines of $S$.
In a sense, $S$ is $t$-independent.
\item Commutativity of matrices $\eta^{(k)}$ reflects a hidden {\bf integrable structure}
of the problem.
\item Within this paradigm, other linear combinations of the same matrices would mean extra time-variables. They can be obtained within consideration of other evolution families. In particular, one pick up an arbitrary knot and consider an evolution family of knots with two parallel pretzel fingers that it induces.
\end{itemize}

\subsection{Representation $R=[3]$}

Let us demonstrate how this scheme works in the case of the second symmetric representation $[3]$. In this case,
\be
\mathfrak{h}_{YZ} = q^{12}A^6\cdot \frac{\sqrt{d_Yd_Z}}{d_{[3]}}\times
\ee

{\footnotesize
\be
\hspace{-1cm}\times\left(\begin{array}{c|cc}
& [33] & [42] \\
\hline
&&\\
\l[33] & 1-\frac{[3][4]\cdot t}{q^4[N-1]} +\frac{[2][3]^2[4]\cdot t^2}{q^7 [N][N-1]}-
\frac{[2]^2[3]^2[4]\cdot t^3}{q^9[N-1][N][N+1]}
& 1-\frac{[2][5]\cdot t}{q^4[N-1]} +\frac{[2][4][5]\cdot t^2}{q^7[N][N-1]}
\\
&&\\
\l[42] & 1-\frac{[2][5]\cdot t}{q^4[N-1]} +\frac{[2][4][5]\cdot t^2}{q^7[N][N-1]}
& 1- \frac{[2]^2 \cdot t}{q^4[4][N+3]} - \frac{[2]^2[5]^2\cdot t}{q^4[3][4][N-1]}
+\frac{[2][5]\big([6][N+3]+2[2]^2[N]-[2][N-1]\big)\cdot t^2}{q^7[3][N-1][N][N+3]} - \frac{[2]^2 [4][5] \cdot t^3}{q^9[N+3][N ][N-1]}
 \\
&&\\
\l[51] & 1-\frac{[6]\cdot t}{q^4[N-1]}
&1- \frac{[2]^2[3]\cdot t}{q^4[4][N+3]} - \frac{[2][5][6]\cdot t}{q^4[3][4][N-1]}
 +\frac{[2]^4[6] \cdot t^2}{q^7 [3][N+3][N-1]}
\\
&&\\
\l[6] & 1 & 1-\frac{[2][3]\cdot t}{q^4[N+3]}
 \\
&&\\
\end{array}\right.\nn
\ee
\be
\hspace{-1cm}\left.
\begin{array}{c|cc}
& [51] & [6] \\
\hline
&&\\
\l[33]
& 1-\frac{[6]\cdot t}{q^4[N-1]}  & 1 \\
&&\\
\l[42]
& 1- \frac{[2]^2[3]\cdot t}{q^4[4][N+3]} - \frac{[2][5][6]\cdot t}{q^4[3][4][N-1]}
 +\frac{[2]^4[6] \cdot t^2}{q^7 [3][N+3][N-1]}
& 1-\frac{[2][3]\cdot t}{q^4[N+3]} \\
&&\\
\l[51]
 & 1- \frac{[2]^2[3]^2\cdot t}{q^4[4][N+3]} - \frac{[6]^2\cdot t}{q^4[3][4][N-1]}
  + \frac{[2][3]^2[4]\cdot t^2}{q^7[5][N+3][N+4]} + \frac{[2]^3[6]^2\cdot t^2}{q^7[5][N-1][N+3]}  - \frac{[2][3] [4][6] \cdot t^3}{q^9[N+4][N+3 ][N-1]}
& 1-\frac{[2][3]^2\cdot t}{q^4[N+3]} + \frac{[2][3]^2[4]\cdot t^2}{q^7[N+3][N+4]}\\
&&\\
\l[6]
& 1-\frac{[2][3]^2\cdot t}{q^4[N+3]} + \frac{[2][3]^2[4]\cdot t^2}{q^7[N+3][N+4]}
& 1-\frac{[3]^2[4]\cdot t}{q^4[N+3]} +\frac{[2][3]^2[4][5]\cdot t^2}{q^7 [N+3][N+4]}-
\frac{[2][3][4][5][6]\cdot t^3}{q^9[N+3][N+4][N+5]} \\
&&\\
\end{array}\right)\nn
\ee
}

\bigskip

\be
= q^{12}A^6\left( \frac{\sqrt{d_Yd_Z}}{d_{[3]}}  -t\cdot \frac{[3]}{q^4}\cdot  \eta^{(1)}
+ t^2\cdot\frac{[2]^2[3]}{q^7}\cdot  \eta^{(2)}
-t^3\cdot \frac{[2][3]}{q^9} \cdot \delta_{YZ}\right)
\ee

In order to obtain this decomposition we proceed in the following way:
\begin{itemize}
\item Derive the evolution matrices (for example, from the 3-strand braid calculation).
\item Divide them by dimensions in order to obtain the peculiar polynomial in $t$,
its coefficients have a clear structure, but still are not fully specified
(for example, the $t$-linear term is a linear combination of terms $[N+\alpha]^{-1}$ with
$\alpha$ from a given set $Y\cup Z/R$, but with yet unspecified $N$-independent coefficients).
\item For some matrix elements  $\mathfrak{h}_{YZ}$, however, the decomposition is unambiguous,
for example, for the first and last lines, and this provides {\it some} information.
\item Now require that $\eta^{(1)}$ has rank two, moreover, one of the eigenvectors is
$\mu_Z=\sqrt{d_Z}{d_R}$, i.e.
$\eta^{(1)}_{YZ} = \alpha \mu_Y\mu_Z + \beta \nu_Y\nu_Z$ with some new unit vector $\nu_Z$.
Since some of the elements of $\eta^{(1)}$ are already known from the previous step,
one can find $\nu_Z$ and the $\eta^{(1)}$-eigenvalues $\alpha$ and $\beta$
(moreover, this system is already overdefined and provides an additional test).
Thus, one knows $\eta^{(1)}$ for all $Y$ and $Z$.
\item With known $\eta^{(1)}$, some new matrix elements acquire unambiguous decomposition.
This provides enough constraints (in fact, again more than enough) to find the next unit
vector $\rho_Z$ contributing, together with the already known $\mu_Z$ and $\nu_Z$, to
$\eta^{(2)}_{YZ} = \alpha^{(2)}\mu_Y\mu_Z + \beta^{(2)} \nu_Y\nu_Z+ \gamma^{(2)}\rho_Y\rho_Z$.
\item Continuing this procedure, one reconstructs step-by-step the entire $\eta$-decomposition
of $\mathfrak{h}$.
All matrices  $\eta^{(k)}$ commute by construction, the non-trivial part of the story
is that such a decomposition {\it exists}.
\end{itemize}

Coming back to our case of $R=[3]$, we get four orthonormal vectors:

{\tiny
\be
\mu_1 = \frac{\sqrt{d_{[33]}}}{d_{[3]}} = \sqrt{[N-1]\over [4][N+2]}, \ \
\mu_2 = \frac{\sqrt{d_{[42]}}}{d_{[3]}}= [3]\sqrt{[N-1][N+3]\over [4][5][N+1][N+2]}, \ \
\mu_3 = \frac{\sqrt{d_{[51]}}}{d_{[3]}}=\sqrt{[2][3][N-1][N+3][N+4]\over [4][6][N][N+1][N+2]}, \ \
\mu_4 = \frac{\sqrt{d_{[6]}}}{d_{[3]}}=\sqrt{[2][3][N+3][N+4][N+5]\over [4][5][6][N][N+1][N+2]}, \nn\\
\nu_1 =  \sqrt{\frac{[N+1]}{[4][N+2]}}, \ \
\nu_2 = \frac{[2][N+4]-[N-1]}{\sqrt{[4][5][N+2][N+3]}},\ \
\nu_3 = \Big([N+3]-[2][N]\Big)\sqrt{\frac{[2][3][N+4]}{[4][6][N][N+2][N+3]}}, \ \
\nu_4 = -[3]\sqrt{\frac{[2][3][N-1][N+4][N+5]}{[4][5][6][N][N+2][N+3]}}, \nn \\
\!\!\!\!\!\!\!\!\!\!\!\!\!\!\!\!\!\!
\rho_1 = -\sqrt{\frac{[N+3]}{[4][N+2]}}, \ \
\rho_2 = \frac{[2][N]-[N+5]}{\sqrt{[4][5][N+1][N+2]}},\ \
\rho_3 = [N]\Big([2][N+4]-[N+1]\Big)\sqrt{\frac{[2][3]}{[4][6][N+4][N+2][N+1][N]}},\ \
\rho_4 = -[3]\sqrt{\frac{[2][3][N+5][N][N-1]}{[4][5][6][N+1][N+2][N+4]}}, \nn \\
\tau_1 = \sqrt{\frac{[N+5]}{[4][N+2]}}, \ \
\tau_2 = -[3]\sqrt{\frac{[N+5] [N+1]}{[4][5][N+3][N+2]}}, \ \
\tau_3 = \sqrt{\frac{[2][3][N][N+1][N+5]}{[4][6][N+2][N+3][N+4]}}, \ \
\tau_4 = -[N+1]\sqrt{\frac{[2][3][N][N-1]}{[4][5][6][N+2][N+3][N+2][N+4]}}
\nn
\ee
}
which form the lines of the matrix $S^{[3]}$,
\be
S^{[3]} = \left(\begin{array}{c|cccc}
& [33] & [42] & [51] & [6] \\ \hline
\emptyset & \mu_1 & \mu_2 &\mu_3 & \mu_4 \\
\hbox{Adj}=[2,1^{N-1}] & \nu_1 & \nu_2 & \nu_3 & \nu_4 \\
\l[2,2,1^{N-2}] & \rho_1 & \rho_2 & \rho_3 & \rho_4 \\
\l[2,2,2,1^{N-3} & \tau_1 & \tau_2 & \tau_3 & \tau_4
\end{array}\right)
\ee
and two non-trivial $\eta$-matrices of ranks two and three:
\be
\eta^{(1)}_{YZ} = \frac{[3]}{[N]}\cdot \mu_Y\mu_Z + \frac{[N+3]}{[N][N+1]} \cdot  \nu_Y\nu_Z \nn \\
\eta^{(2)}_{YZ} = \frac{[3]}{[N][N+1]}\cdot \mu_Y\mu_Z
 + \frac{[2][N+3]}{[N][N+1][N+2]} \cdot  \nu_Y\nu_Z + \frac{[N+4]}{[N][N+1][N+2]} \cdot \rho_Y\rho_Z
\ee
in addition to the rank one matrix $\eta^{(0)}_{YZ} = d_{[3]}\cdot \mu_Y\mu_Z = \frac{\sqrt{d_Yd_Z}}{d_{[3]}}$
and the rank four matrix $\delta_{YZ} = \mu_Y\mu_Z + \nu_Y\nu_Z + \rho_Y\rho_Z + \tau_Y\tau_Z$. The first index of $S^{[3]}$ runs over the representations in the decomposition $[3]\otimes\bar [3]=\emptyset\oplus\hbox{Adj}\oplus [2,2,1^{N-2}]\oplus [2,2,2,1^{N-3}]$.

The other exclusive Racah matrix $\bar S$ is obtained from $S$ by the rule (\ref{bSfromS}).

\subsection{Direct and inverse problems for $\eta$-matrices}

To clarify the notion of $\eta$-matrices a little more, we repeat once again
the expression (\ref{12}) of the evolution matrix $\mathfrak{h}$ through the Racah matrix $S$
(assuming the eigenvectors are its lines):
\be
\mathfrak{h}_{YZ} = q^{4\nu_R}A^{2|R|}\cdot
\underbrace{d_R \cdot \overbrace{S_{Y1}S_{Z1}}^{\mu_Y\mu_Z}}_{\frac{\sqrt{d_Yd_Z}}{d_R}}
\left( \frac{K_1}{q^{4\nu_R}A^{2|R|}d_R}\,{\cal E} +
\sum_{k\neq 1}\frac{K_k}{q^{4\nu_R}A^{2|R|}d_R}\cdot \frac{S_{Yk}S_{Zk}}{S_{Y1}S_{Z1}} \right)
\ee
where ${\cal E}$ is a matrix with all unit matrix elements,
$K_k$ are expanded in $t$, and the coefficient in front of $(-t)^k$
multiplied by $d_R S_{Y1}S_{Z1}$ defines $\eta^{(k)}$.
As was already mentioned, it is applicable not only to the pretzel family $Pr(m,n,\pm  2)$,
but actually to any family with two pretzel fingers attached to anything else.

If the Racah matrix $S$ is already know, the only
point is that for $A=q^N$ the evolution matrix  $\mathfrak{h}$ is {\it not} fully expressed
through the ($N$-dependent) quantum numbers, and explicit dependence on $\{q\}$ is encoded
in the form of the $t$-expansion, $t = -A^{-1}\{q\}$.
As to ingredients of the above formula, the quantum dimensions $d$ and
matrix elements $S_{Yk}$ are made from quantum numbers, only the eigenvalues $K_k$
and the framing pre-factors are {\it not}.
Therefore, in a good sense, the $t$-expansion with $\eta$-matrices can be considered
as reflecting the differential expansion for the {\it complement} of the two pretzel fingers:
for the family $Pr(m,n,\pm \bar 2)$ this is literally the expansion of the finger
$K_Z = (\bar S\bar T^{\pm 2} \bar S)_{\emptyset X}$.

In the simplest example of $R=[1]$, the two eigenvalues are
\be
\frac{K_1}{A^2} &=& \frac{1}{A^2}\cdot K^{[1]}_\emptyset \ = \
\frac{1}{A\{q\}}\Big(A\{A\}-\{q\}^2\Big) = d_{[1]} - t\nn \\
\frac{K_2}{A^2} &=& A^{-2}K^{[1]}_{\rm Adj}  \ = \
-\frac{A\{q\}}{A^2}=-t
\ee
and
\be
\mathfrak{h}^{[1]}_{YZ} = A^2 \cdot d_{[1]}\cdot S_{Y1}S_{Z1}\left\{
\underbrace{\frac{K_1}{A^2d_{[1]}}}_{1-\frac{t}{d_{[1]}}}
\left(\begin{array}{cc} 1&1\\1&1 \end{array}\right) +
\underbrace{\frac{K_2}{A^2d_{[1]}}}_{-\frac{t}{d_{[1]}}}
\left(\begin{array}{cc} \sqrt{\frac{[N-1]}{[N+1]}} & -1 \\
-1 & \sqrt{\frac{[N-1]}{[N+1]}}\end{array}\right)\right\}
=\nn\\
= A^2 \cdot d_{[1]}\cdot S_{Y1}S_{Z1}\left(\begin{array}{cc}
1- \frac{[2]}{[N+1]}\cdot t & 0 \\ 0 & 1- \frac{[2]}{[N+1]}\cdot t
\end{array}\right)
\ee
Likewise, for $R=[2]$  the three eigenvalues are given by (\ref{ev2}),
and they need be divided by the framing factor $q^4A^4$.
After that, the {\bf third} eigenvalue, $K^{[2]}_{[2,2,1^{N-2}]}$ is immediately equal to $[2]t^2q^{-3}$.
The {\bf second} one becomes
\be
\frac{1}{q^4A^4}K^{[2]}_{\rm Adj} = -\frac{1}{A^2}
\Big( q^{-m}A\underbrace{\{Aq^m\}}_{[N+m]\{q\}} + \underline{q^{-2m} - q^{-6}+q^{-2}-1}\Big)
\ee
The first term here is a quantum number multiplied by $tq^{-m}$,
and $m$ should be chosen so that the underlined combination gets expressible through $t$.
This means that it should be proportional to $\{q\}^2$, and for this we should take $m=2$.
Thus, we {\it deduce} the decomposition
$ - \frac{t}{q^2} [N+2] + \frac{[2]t^2}{q^3}$ familiar from (\ref{ev2rev}).
Similarly, for the {\bf first} eigenvalue we want the following form:
\be
\frac{1}{q^4A^4}K^{[2]}_{\emptyset} = d_{[2]} - \beta\cdot t\cdot \frac{\{Aq^m\}}{\{q\}} + \gamma\cdot t^2
\ee
which means that, after multiplication by $\{q\}^2$, one gets a decomposition of the polynomial,
where the coefficients $\beta$ and $\gamma$ are proportional to $\{q\}^2$ and $\{q\}^4$
respectively (rather than to $\{q\}$ and $\{q\}^2$).
These additional powers impose conditions on $m$, $\alpha$ and $\beta$ (actually, an
overdefined set of conditions), and the solution for the decomposition problem is
\be
K^{[2]}_{\emptyset} = q^4A^4\left(d_{[2]} - \frac{t}{q^2} [2][N+1] + \frac{[2]t^2}{q^3}\right)
\ee
again in accordance with (\ref{ev2rev}).
For generic symmetric representation $R=[r]$, the eigenvalues
\be
K^{[r]} = q^{2r(r-1)}A^{2r} \sum_{i=0}^r  C_i t^iq^{-i(4r-3-i)/2}
\ee
where $C_i$ are coefficients made from quantum numbers, $2r(r-1) = 4\nu_{[r]}$ and, in the term with $i=r$, the power of $q$ is actually $-3\nu_{[r]}$.

\bigskip

For $R=[22]$, one naively gets:
\be
\frac{K_1}{A^8} &=& \frac{K_\emptyset^{[22]}}{A^8}=\underbrace{\frac{[N-1][N]^2[N+1]}{[2]^2[3]}}_{d_{[22]}} -
\frac{[2]^2[N-1][N][N+1]}{[3]}\cdot t + 2[3][N]^2\cdot t^2 - ([2]^2+1)[3][N]\cdot t^3
+ [2]^2[3]\cdot t^4\nn\\
\frac{K_2}{A^8} &=&\frac{K_{[2,1^{N-2}]}^{[22]}}{A^8}= -\frac{[N-2][N][N+2]}{[3]}\cdot t + [3][N]^2\cdot t^2 - [2]^2[3][N]\cdot t^3
+  [2]^2[3]\cdot t^4 \nn \\
\frac{K_3}{A^8} &=&\frac{K_{[2,2,1^{N-4}]}^{[22]}}{A^8}= \boxed{q^{-1}[N+1][N+2]\cdot t^2} - [2][3][N+1]\cdot t^3 + [2]^2[3]\cdot t^4 \nn \\
\frac{K_4}{A^8} &=&\frac{K_{[4,2^{N-2}]}^{[22]}}{A^8}=   \boxed{q[N-1][N-2]\cdot t^2} - [2][3][N-1]\cdot t^3 + [2]^2[3]\cdot t^4
\label{ev22deco}\\
\frac{K_5}{A^8} &=&\frac{K_{[4,3,2^{N-4},1]}^{[22]}}{A^8}= -[3][N]\cdot t^3 +[2]^2[3]\cdot t^4 \nn \\
\frac{K_6}{A^8} &=&\frac{K_{[4,4,2^{N-4}]}^{[22]}}{A^8}= [2]^2[3]\cdot t^4 \nn
\ee
For the transposition invariant diagrams $R$, however, there should be no bare
powers of $q$ in the $t$-expansions, only quantum numbers,
while in (\ref{ev22deco})  there still {\it are}, in  the two boxed terms.
They are eliminated by the substitutions
\be
q = \frac{[N+2]-[N]+t}{[N+1]-[N-1]}, \ \ \ \ \  q^{-1} = \frac{[N]-[N-2]-t}{[N+1]-[N-1]}
\ee
which contain $t$ and, thus, change the decompositions of two eigenvalues for
\be
\frac{K_3}{A^8} = [N+1][N+2]\frac{[N]-[N-2]}{[N+1]-[N-1]}\cdot t^2 - [N+1]\left(\frac{[N+2]}{[N+1]-[N-1]}+[2][3]\right)\cdot t^3 + [2]^2[3]\cdot t^4 \nn \\
\frac{K_4}{A^8} = [N-1][N-2]\frac{[N+2]-[N]}{[N+1]-[N-1]}\cdot t^2 + [N-1]\left(\frac{[N-2]}{[N+1]-[N-1]}-[2][3]\right)\cdot t^3 + [2]^2[3]\cdot t^4
\ee

\bigskip

Thus, if the Racah matrix and the eigenvalues are already known, one can easily reconstruct
the evolution matrix $\mathfrak{h}$ and its $\eta$-decomposition:
this is just a simple application of the arborescent knot calculus from
\cite{MMMRS,MMMRSS}.
The point of the present paper was the use of this knowledge for solving the
{\it inverse} problem: reconstruction of $S$ from known $\mathfrak{h}$.
We explained that formally this is a straightforward linear algebra problem,
but actual diagonalization of complicated matrices with entries that contain
square roots, is a nearly un-doable by MAPLE and Mathematica, hence, one needs
tricks to do it.
There are many of them, from an analytical continuation from numeric values
of parameters and to explicit use of Cramer's rule.
Knowledge of additional structures like $\eta$-decomposition
provides additional technical advantages, and is also of certain conceptual value.
One of the hopes is that it can be used to clarify the situation with {\it differential}
\cite{DGR,IMMMfe,evo,GGS,diffarth,Kondef}, perturbative (Vassiliev) \cite{Vass}
and genus (Hurwitz) \cite{MMS} expansions,
to which it is clearly related.

It deserves noting that the $t$-decomposition of the simple pretzel finger $F(\bar 2)$,
i.e. of concrete eigenvalues that we studied in this subsection,
is a simple and straightforward part of the {\it inverse} problem,
and it is exactly the fact that it is easily solvable which makes possible the initial
step in the algorithm of sec.\ref{alg}.

\section{Conclusion}

To conclude, we have made yet another step in evaluating the colored HOMFLY polynomials for arbitrary arborescent
knots, which can be now extended in many directions.

One should extend the tables of \cite{knotebook} to include these new colored polynomials
for all arborescent knots in the Rolfsen list.
This can be done with the help of the powerful {\it families method} of \cite{MMfam,MMMRSS}.

One also should apply the technique developed in this paper to calculating the exclusive $[3,1]$ Racah matrices in order to complete evaluating the colored HOMFLY polynomials in all representations of size $|R|\leq 4$: the inclusive Racah matrices in this case have been calculated in \cite{MMMS31}, which allowed us to evaluate the 3-strand braid polynomials, those for the arborescent knots still remains unavailable.

Another immediate thing to do is to search for $\hat A$-polynomial equations \cite{Gar} (see also a review in \cite{MMeqs}),
differential expansions \cite{diffarth,Kondef}, hyper- \cite{AS,Che,GGS} and super-
\cite{DGR,DMMSS} polynomials, to learn more about factorization properties \cite{Konfact},
and about the Vassiliev \cite{Vass} and Hurwitz expansions \cite{MMS} extending the sample analysis
in \cite{MMMS22,MMMS31} from 3-strand braids to arborescent knots.

Also a new breath is now given to the Racah calculus, where, first, one can attack the next
principal barrier of $R=[4,2]$ and, second, proceed to composite mixing matrices
needed to handle knots with more than $3$ strands.

Conceptually, the most interesting fact is that the Racah matrices look not the {\it elementary}
({\it primary}) objects
in the theory, instead they can be {\it derived} from something else.
In the context of the present paper, $S$ are diagonalizing matrices of the evolution coefficients
matrices $h$, which in their turn are averages of even more elementary
coefficients for the 4-parametric evolution.
However, as demonstrated in \cite{MMMS31}, even these coefficients look like {\it composites}:
sums of different items.
The true elementary fully factorizable objects still remain to be identified.
For a related  recent suggestion see \cite{GLL}.

\section*{Acknowledgements}

This work was funded by the Russian Science Foundation (Grant No.16-12-10344).

\end{document}